\documentclass{article}

\usepackage{titling}
\title{Does the Internet deserve everybody?}

\usepackage[sc]{mathpazo} 
\usepackage[T1]{fontenc} 
\usepackage{microtype} 
\usepackage[hmarginratio=1:1,top=32mm,columnsep=20pt]{geometry} 
\usepackage[hang, small,labelfont=bf,up,textfont=it,up]{caption} 
\usepackage{booktabs} 
\usepackage{float} 
\usepackage{paralist} 

\usepackage{abstract} 

\usepackage{titlesec} 
\renewcommand\thesubsection{\Roman{subsection}} 
\titleformat{\section}[block]{\large\scshape\centering}{\thesection.}{1em}{} 
\titleformat{\subsection}[block]{\large}{\thesubsection.}{1em}{} 

\usepackage{fancyhdr} 
\usepackage[pdftex,
            pdfauthor={Yehia Elkhatib, Gareth Tyson, and Arjuna Sathiaseelan},
            pdftitle={Does the Internet deserve everybody?},
            pdfsubject={Ethicality of Internet initiatives in developing regions},
            pdfkeywords={Internet, GAIA, developing regions}]{hyperref}

\usepackage{textcomp} 

\title{\fontsize{24pt}{10pt}\selectfont\textbf{\thetitle}}

\author{
\large
\textsc{Yehia Elkhatib\textsuperscript{$\bullet$}, Gareth Tyson$^{\diamond}$, Arjuna Sathiaseelan\textsuperscript{\textdaggerdbl}}\\[2mm] 
\normalsize \textsuperscript{$\bullet$}Lancaster University  $^{\diamond}$Queen Mary, University of London  \textsuperscript{\textdaggerdbl}University of Cambridge\\ 
\normalsize \href{mailto:y.elkhatib@lancaster.ac.uk}{y.elkhatib@lancaster.ac.uk} 
}
\date{This is a pre-print. The final version is made available by ACM at:\\
 \url{http://dx.doi.org/10.1145/2793013.2793018}}

\begin{document}

\maketitle

\thispagestyle{fancy} 

\begin{abstract}
There has been a long standing tradition amongst developed nations of influencing, both directly and indirectly, the activities of developing economies. Behind this is one of a range of aims: building/improving living standards, bettering the social status of recipient communities, etc. In some cases, this has resulted in prosperous relations, yet often this has been seen as the exploitation of a power position or a veneer for other activities (e.g.\ to tap into new emerging markets). In this paper, we explore whether initiatives to improve Internet connectivity in developing regions are always ethical. We draw a list of issues that would aid in formulating Internet initiatives that are ethical, effective, and sustainable.
\end{abstract}

\section{Introduction}
\label{sec:intro}

Is the expansion of the Internet an inherently noble goal? We have increasingly begun to witness the promotion of Internet deployment as a means to social betterment. To date, only 37.9\% of the world's population have Internet access~\cite{itud2013}. Unsurprisingly, the Western world is disproportionately connected, with 75.7\% of the developed world having access to the Internet in 2013, as opposed to only 29.9\% in developing countries~\cite{itud2013}. Economic benefits, such as GDP, are also disproportionately biased, with developed economies ahead by a factor of about 25\%~\cite{globaltrends2013}. This observation has led many countries and organisations to actively engage with developing nations to assist them in improving their information and communication technologies (ICT) capabilities (e.g.\ Internet infrastructure and data centres) as a means of improving their economic standing.

On the one hand, this could be seen as an act of altruism to improve the living standards and social status in the recipient community. However, the motives may not be entirely selfless. There are a variety of reasons why a highly developed country or organisation might assist a less developed one. A clear example is the incentive to foster long-term economic relationships. Developing and emerging economies (DEEs)\footnote{Both are loosely defined terms referring to countries with low GDP. A distinction made between them is that emerging economies enjoy relatively more stable governments and a growing consumer market with disposable income.} represent a huge financial potential in terms of both natural resources and trading partners. Early investment in such countries (e.g.\ Kazakhstan, Nigeria, Gulf states) could mean significant long-term gains for the stakeholders involved (post liberalised India being a great example). We have already seen such strategies employed by several prominent multi-national technology companies such as Microsoft, Facebook, Google, and others. 

Initially, one might consider the motivation of such actions irrelevant, arguing, instead, that developing ICT in a country is inherently good. Whereas this is a justifiable argument, we pose a question: Is Western intervention in developing countries' ICT infrastructures always ethical? This raises many further questions that could have far reaching implications on the geo-political climate for decades to come. Maybe the West's faith in their strategies could create short, medium or even long-term damage.
This paper explores some of these questions and outlines a few key issues that need to be addressed by any ICT development project in order to be ethical, effective, and sustainable.

\section{Let them have Internet}
\label{sec:for}

The Internet has undeniably been a benefit to millions of people. The free flow of information has produced economic growth, ubiquitous access to information, improved business efficiency, etc. It is estimated that the Digital Economy boosted the world's economy by \$193 billon and created 6 million jobs in 2011~\cite{globaltrends2013}. Thus, many works have argued that Internet penetration is correlated with social and economic development (cf.~\cite{un2013report}), with others going as far as to claim that the former is an enabler for the growth of the latter~\cite{manyika2011great,amiri2013internet}. The Internet has arguably helped topple dictators, better health interventions, reduce poverty, as well as enable near-impossible achievements such as monitoring lynch-logging in the Amazonian rainforest through Google Earth and replanting accordingly~\cite{ragogmakan}. It therefore seems intuitive that improving Internet connectivity in developing regions would engender various benefits.

\section{But do they really want it?}
\label{sec:against}

Many of those who are advocating the universal benefits of the Internet are those who have \emph{personally} benefited (or have the potential to benefit). For example, the service provider who would expand unopposed into a new market or the network operator who would lay down hundreds of cable miles. 

Some may argue that this is irrelevant, as long as it better enables Internet access. However, an important question here is whether that is essentially better or simply an effort to measure development ``using the industrialised countries as a yardstick''~\cite{wilson2002understanding}. In other words, are such benefits felt by the wider population or just these leading players? Within such an argument, it becomes implicit that the Internet brings both winners and losers. Many companies have closed with the advent of Internet shopping (e.g.\ it is estimated that the growth of online shopping in the UK could lead to the closure of one in five high street stores by 2018~\cite{guardian}). The proliferation of illegal goods, child pornography, and online radicalisation highlight just some of the societal dangers. 

Naturally, any technology comes with its own mix of advantages and disadvantages. However, whereas populaces in the West have had a slow and steady introduction to these risks, it is likely that people in developing regions would be ``thrown in at the deep end''. In the words of Uncle Ben from Spiderman, with great power comes great responsibility; exposing whole communities to the Internet without appropriate levels of introductory education could be a highly risky strategy, creating significant disruption (cf.~\cite{Schuppan2009egov}). For instance, an obvious risk of Internet deployment in an country with an authoritarian regime would be the ability for unregulated Internet surveillance. Without appropriate education, civilians could open themselves to many attacks by exposing browsing habits, such as those relating to political activism and journalism, or sexual content that is deemed taboo by a certain society. More mundane, but still worrying, are the threats of malware, spam, online fraud, etc.

\section{How to fail}
\label{sec:fail}

The developed world has had many opportunities to learn the problems facing ICT Development projects. We therefore know exactly what steps to take to \emph{fail}. Perhaps most prominently, too many projects do not state (in detail) how the improvement of ICT capabilities will improve the lives of the target users, or who these users really are. These projects are reduced to efforts of blindly emulating the technological developments in the West, and, in turn, they reduce the recipient user community to a simple homogeneous group~\cite{wilson2002understanding}. This is not the case. Users are diverse, with many different needs, and these need to be addressed and communicated within any development project.

Another fundamental issue seldom addressed is whether or not the target users actually \emph{want} the Internet at all. There is of course an argument that they do not know what they're missing, and that the Internet could radically improve their lives. However, in many cases, it is actually extremely difficult to persuade the target community to become users~\cite{Cecchini2004egov}. In other cases, the community just wants the well-being that the Internet could bring, without the Internet itself. A recent example is that of policies relating to irrigation being disseminated to farmers and local authorities in rural Egypt. Elaborate stakeholder engagement concluded that this is best done via printed guides rather than the Internet, as it was found that the latter just forms a literacy and accessibility hurdle~\cite{amr15refguide}. In other words, citizens and local officials wanted on-demand access to information but not via the Internet.

This is not to say that the projects that fail are not well meaning though. Many are inspirational. We frequently see ambitious goals, such as: 
giving communities in remote rural areas quick access to medical expertise for urgent diagnosis and advice; or 
educating school children in economically impoverished areas to gain knowledge they would otherwise not get. Although these goals are admirable, a well-trodden failure route is to not associate such goals with strategies in the recipient community/nation, e.g.\ adult literacy initiatives.
Such organisational alignment has been found to be more important than financial funding~\cite{Chinn2010ict,Weerakkody2011exploring}. 
Another frequent issue is that these goals are not mapped into long-term implications and sustainability plans. Of course, it is often difficult to plan for the long-term beyond the end of the initial pilot. Yet, it has to be noted that a majority of ICT for Development projects have failed, either partially or totally, in achieving their stated objectives because of poor sustainability planning~\cite{best2002rural,heeks2002dev,heeks2003most,Avgerou2008isdc}. This is evident from the sustainability failure of many rural telecenters, e.g.\ the E-Srilanka programme funded by the World Bank in 2004, SARI (Sustainable Access in Rural India)~\cite{BestKumar2008sari} project etc.

The most obvious sustainability plan involves switching an ICT infrastructure into a (local) commercial footing~\cite{Kuriyan2008ict4dev}, a strategy that comes with its own risks. Although such a transformation might be positive in terms of sustaining the infrastructure, it is important to ensure that the initial societal goals are not undermined by such a change. 

A question arises here of whether deploying an infrastructure which is subsequently disabled due to lack of funding might be more damaging than not providing it at all. For instance, a community that has become dependent on a technology (e.g.\ messaging service) could suffer if it is subsequently removed without replacement. 

Finally, we wish to highlight an oft cited concern (particularly in some popular media outlets). It is undeniable that many areas targeted for development have issues with corruption of varying severities. Sadly, there are many examples of project resources being drained away to individuals and private firms once strict enforcement of resource allocation is removed. Beyond the economic dimension, infrastructures could be used in ways they were not intended to such as spying on and prosecuting political activists. Of course, this is a challenging concern as it must be addressed on many different levels, most of which cannot be enforced by the development project itself. Instead, it becomes necessary to shape all policy steps with the focus on mitigating the impact of potential corruption: a process which, itself, can undermine the success of the project.

\section{Sanity Checklist}

From the above, it is clear that achieving constructive, long-term, ethical Internet deployment initiates is non-trivial. Thus, we next offer a checklist of considerations that any future initiates should consider. We do not intend this to be exhaustive, and many issues overlap, however, common to all of them is the value of \emph{transparency}.

\subsection{Trust}
As already discussed, many Internet initiatives require significant capital expenditures and expertise to set up.
This places huge (i.e.\ scary) amounts of power in the hands of politicians.
For this to be ethical and to establish trust within the recipient community, especially in politically turbulent and corrupt environments, the following questions should clearly answered to the public:
\begin{itemize}
 \item Who are the stakeholders involved, both internally and externally? What is the exact role of each of the stakeholders?
 \item What is the management timeline; i.e.\ what are the targets for the different phases of the project (e.g.\ foundation, pilot, public launch, etc.)? And who is managing these?
 \item Who are the decision makers in the initiative? How much control do they have? Who are they answerable to? What are the mechanisms for members of the local community to participate in the decision making process?
\end{itemize}
Answering these questions is vital in order to prescribe local management without being viewed internally as a political tool, or risk being rejected or demonised. 
It is important to also note that to attain trust and transparency, the project should employ varied dissemination channels that are accessible to the majority, if not all, of the target population. 
For instance, an initiative targeted at helping a farming community should communicate through village community leaders, farmers syndicates/associations, local agricultural authorities, and schools. This is related to the final point in the following subsection.

\subsection{Economic Sustainability}
Also related to transparency, the economic model attached to the initiative should be made available and accessible to the public. This includes the timetable of tangible targets, whether commercial or otherwise, and funding/financing influx and contingency plans. This should make it easy to ascertain whether the initiative is economically affordable and self-sustainable, or a ``white elephant'' that would need constant pumping of funds. In other words, an exit strategy is desired, allowing a community or nation to become independently responsible for the initiative.

The plan should also clearly describe the benefactors and beneficiaries, as well as maintenance funds and their channels and overseers. This should help identify the economical potential of the project from the perspective of the different stakeholders, and whether there is potential for the the project to degenerate to serve alternative goals shortly after deployment. Optimally, there should be some distinction between such financial machinery and the management organisation to avoid potential conflict of interests and to better manage coordination with different stakeholders~\cite{BestKumar2008sari}.

Finally, the economic and recruitment plans should ideally be broad enough to include partnerships with both public and private sectors. This should provide opportunities to stimulate socioeconomic and educational reforms, and to engage with a wider cross-section of society.

\subsection{Impact}

As we have discussed, it is not sufficient to claim that introducing Internet connectivity or achieving certain ICT metrics will magically transform the recipient community for the better. 
Instead, the initiative should be expected to clearly identify pathways to benefit. 
Otherwise, there is a risk for the initiative to be perceived as opportunistic (big business breaking into a new market) or exploitative (by corrupt leadership). 
Hence, a study carried out in liaison with community representatives should analyse the effect on the local community, including but not limited to:
\begin{itemize}
 \item How does the initiative empower the people? What opportunities does it provide: educational, organisational, commercial, social, etc.?
 \item What sort of business/social links are expected to be built and with whom? What is the expected effect on the political structure?
 \item How would the initiative affect the dynamics of the local community? Will it provide a platform for certain groups over others?
 \item How will users be recruited? How do we enable successful adoption? How do we measure successful adoption?
\end{itemize}

It is also important to acknowledge that the initiative builds a two-way bridge that expands the international online community. It is both myopic and disrespectful 
to assume, like many initiatives seem to do, that it is only communities in DEEs that stand to benefit from such initiatives. 
An ethical initiative would consider how to enrich the international community from the newly arrived at resources of local knowledge and experience in different contexts, as their constraints usually spawn innovative use of ICT capabilities.

\subsection{Community Capacity Building}
A commitment to being connected and joining an international community requires continuous learning processes, not just setting up physical infrastructures and software systems. 
It is a responsibility to ensure that new users introduced to the Internet are properly educated about the means and potential risks of it. 
This is a shared responsibility. Those who facilitate an initiative from outside should also provide training processes, educational material, shared best practices and lessons learnt. Needless to say, these should be made available in local languages and with respect to local traditions. Moreover, those who manage or oversee it locally need to include the provision of technical support channels, Internet safety courses, and material for responsible Internet use as a running cost.
           
\section{Discussion and Conclusion}
\label{sec:disc}

The above sections have highlighted some of the many pros and cons that can emerge from deploying Internet connectivity in developing (and emerging) regions. It is important to note that we are \emph{not} against the deployment of ICT infrastructure in developing regions. In fact, we passionately support it. Thus, the key tenet of this paper is to state that the argument is complex and nuanced -- not black and white. Exploring and understanding this nuance is key to sustainable, long-term solutions.

This argument is driven by the fact that developed nations have made mistakes in the past. Interference with developing countries has been commonplace throughout the centuries. Empire building during the 17--19\textsuperscript{th} centuries was often motivated by the so called ``cultural'' contributions made by the occupying state. However, this was often a thinly veiled disguise for exploitation of both people and natural resources. The British Empire perhaps offers one of the most prominent (and egregious) examples. The free-market economy values of Britain, for example, brought the East India Company to India; a profit-driven company backed by the British military. Whilst propaganda suggested that this was a mutually beneficial arrangement, it was, in fact, a mechanism by which Britain could establish economic dominance on the sub-Indian continent. Could developed nations and organisations be driven by such motives in their push for ICT expansion in new emerging and developing economies?

Drawing a generalisable conclusion from these arguments is a near impossible feat. The many complex economic, political and cultural aspects of the countries that fall under the umbrella term ``developing'' make such an accomplishment intractable. However, the clear issues that have arisen during the developed world's past suggest that such conclusion should, at least, be attempted. Thus, we argue here is no one-size-fits-all. A more sophisticated methodology beyond ``let's do it'' should be formed. Cost-benefit analysis is a methodology devised for such purposes. Key to this, however, is transparency: the reasons and mechanisms behind decisions should be announced and understood by the wider population, not confined within governmental buildings. A particularly important goal should be to achieve global access for \emph{all}. As such, cost benefit should not be driven by sole economic factors that may cease deployment once a given cost-revenue threshold has been reached.

Ultimately, being firm about whether advancing the ICT infrastructure is good or bad can only be done within a very specific context. We therefore conclude with questions, rather than answers. In what directions will the ICT be improved? Who are the local and outside stakeholders that will be involved both in the planning, actuation, and follow up development stages? What are the socio-economic repercussions of the said development? What political obstacles, if any would need to be overcome and how?

\section*{Acknowledgments}
Yehia Elkhatib was supported by the CHIST-ERA Dionasys project via EPSRC grant reference EP/M015734/1. Arjuna Sathiaseelan was supported by the European Commission Horizon 2020 RIFE project grant reference 644663.

	\bibliographystyle{abbrv}
	\bibliography{ethics-paper}

\end{document}